\documentclass[12pt]{article}
\usepackage{mathrsfs}
\usepackage{graphicx}
\usepackage{amsmath}
\usepackage{amssymb}
\usepackage{caption2}
\begin{document}
\newcommand  {\ba} {\begin{eqnarray}}
\newcommand  {\be} {\begin{equation}}
\newcommand  {\ea} {\end{eqnarray}}
\newcommand  {\ee} {\end{equation}}
\renewcommand{\thefootnote}{\fnsymbol{footnote}}
\renewcommand{\figurename}{Figure.}
\renewcommand{\captionlabeldelim}{.~}

\vspace*{1cm}
\begin{center}
 {\Large\textbf{A Model of Fermion Masses and Flavor Mixings with Family Symmetry $SU(3)\otimes U(1)$}}

\vspace{1cm}
 \textbf{Wei-Min Yang}* \hspace{0.5cm} \textbf{Qi Wang} \hspace{0.5cm} \textbf{Jin-Jin Zhong}

\vspace{0.4cm}
 \emph{Department of Modern Physics, University of Science and Technology of China, Hefei 230026, P. R. China}

\vspace{0.2cm}
 \emph{*E-mail: wmyang@ustc.edu.cn}
\end{center}

\vspace{1cm}
 \noindent\textbf{Abstract}: The family symmetry $SU(3)\otimes U(1)$ is proposed to solve flavor problems about fermion masses and flavor mixings. It's breaking is implemented by some flavon fields at the high-energy scale. In addition a discrete group $Z_{2}$ is introduced to generate tiny neutrino masses, which is broken by a real singlet scalar field at the middle-energy scale. The low-energy effective theory is elegantly obtained after all of super-heavy fermions are integrated out and decoupling. All the fermion mass matrices are regularly characterized by four fundamental matrices and thirteen parameters. The model can perfectly fit and account for all the current experimental data about the fermion masses and flavor mixings, in particular, it finely predicts the first generation quark masses and the values of $\theta^{\,l}_{13}$ and $J_{CP}^{\,l}$ in neutrino physics. All of the results are promising to be tested in the future experiments.

\vspace{1cm}
 \noindent\textbf{Keywords}: family symmetry; fermion mass; flavor mixing; neutrino physics

\vspace{0.3cm}
 \noindent\textbf{PACS}: 12.10.-g; 12.15.Ff; 14.60.Pq

\newpage
 \noindent\textbf{I. Introduction}

\vspace{0.3cm}
 The precise tests for the electroweak scale physics have established plenty of knowledge about the elementary particles \cite{1}. The standard model (SM) has been evidenced to be indeed a very successful theory at the current energy scale \cite{2}. However, there are some imperfections in the SM, among other things, a ugly defect is too many parameters exist in the Yukawa sector. As a result, fermion masses and flavor mixings seem intricate and ruleless. During the past decade a series of new experiment results about $B$ physics and neutrino physics have told us a great deal of information about flavor physics \cite{3}. What deserves to be paid special attention are some facts as follows. The mass spectrum of quarks and charged leptons emerges a large hierarchy, which ranges from one MeV to a hundred GeV or so \cite{1}. The neutrinos have been verified to have nonzero but Sub-eV masses \cite{4}, nevertheless, that their nature is Majorana or Dirac particle has to be further identified by experiments such as $0\nu\beta\beta$ \cite{5}. On the other hand, the flavor mixing in the quark sector is distinctly different from one in the lepton sector. The former has small mixing angles and its mixing matrix is close to an unit matrix \cite{6}, whereas the latter has bi-large mixing angles and its mixing matrix is close to the tri-bimaximal mixing pattern \cite{7}. In the lepton mixing, it is yet in suspense whether $sin\theta_{13}$ is zero and the $CP$ violation vanishes or not \cite{8}. These impressive puzzles always attract great attention\cite{9}, and also are expected to be explained by new theories beyond the SM. The issues in the flavor physics possibly implicate great significance. They are not only bound up with origin of matter in the universe \cite{10}, but also in connection with the genesis of the matter-antimatter asymmetry and the original nature of the dark matter \cite{11}.

 Any new theory beyond the SM has to be confronted with the various intractable issues mentioned above, however, some approaches and theoretical models have been proposed to solve them \cite{12}. For instance, the Froggatt-Nielsen mechanism with $U(1)$ family symmetry can account for mass hierarchy \cite{13}. The discrete family group $A_{4}$ can lead to the tri-bimaximal mixing structure of the lepton mixing matrix \cite{14}. The non-Abelian continuous group $SU(3)$ is introduced to explain the neutrino mixing \cite{15}. In \cite{16}, the model with the family group $SO(3)$ successfully accommodates the whole experimental data of quarks and leptons. In addition, some models of grand unification (GUT) based on $SO(10)$ symmetry group can also give some reasonable interpretations for fermion masses and flavor mixings \cite{17}. Although these theories seem successful in explaining some of the flavor problems, it seems very difficult for them to solve the whole flavor problems all together. It is especially hard for some models to keep the principle of the smaller number of parameters. On all accounts, it now remains to be a large challenge for theoretical particle physicists to uncover these mysteries of the flavor physics.

 As remarked above, the flavor puzzle is very difficult and complex to solve completely. In this works, we consider a new approach and construct a model with fewer parameters to try to understand these problems. On the one hand, we believe that there is some inherence relations among all kinds of fermion mass and mixing parameters. The family symmetry $SU(3)_{F}$, which is a family symmetry group of the three generation fermions, is appropriate for seeking the relations. In addition, we introduce some super-heavy fermion and flavon fields which appear only at the high-energy scale. They communicate with the low-energy fermions of the SM by flavor interaction. The different super-heavy fermions and flavons are distinguished by an appended Abelian group $U(1)_{N}$. On the other hand, the model with two Higgs doublets and three generation right-handed neutrino singlets is theoretically well-motivated extension of the SM\cite{18}. On that basis, we introduce a real scalar singlet and append a discrete group $Z_{2}$, under which these scalar and right-handed neutrino singlets all reverse sign. The model goes through three steps of breakings. Firstly, the family symmetry $SU(3)_{F}\otimes U(1)_{N}$ is broken at the flavon dynamics scale. It is accomplished by means of the flavon fields developing the vacuum structures along the specific directions. After all the super-heavy fermions are integrated out and decoupling, the low-energy effective theory is elegantly obtained. Secondly, the discrete symmetry $Z_{2}$ is broken at the middle-energy scale by the real scalar developing non-vanishing vacuum expectation value (VEV). This directly causes that the effective Yukawa couplings of neutrinos are drastically suppressed, thus far smaller than ones of the other fermions. This also becomes a source of the neutrino tiny masses. Lastly the electroweak symmetry breaking is completed, all of quarks and leptons attain Dirac masses. All the fermion mass matrices are regularly given and characterized only by the four fundamental matrices and fewer parameters. Finally, the model can naturally and correctly give rise to the fermion mass spectrum and flavor mixing angles. All the numerical results are very well in agreement with the current experimental data.

 The remainder of this paper is organized as follows. In Section II we outline the model. In Sec. III, the symmetry breaking procedures are introduced and the fermion mass matrices are discussed. In Sec. IV, we give the detailed numerical results about the fermion masses and flavor mixings. Sec. V is devoted to conclusions.

\vspace{1cm}
 \noindent\textbf{II. Model}

\vspace{0.3cm}
 The model is based on the symmetry group
 $SU(3)_{C}\otimes SU(2)_{L}\otimes U(1)_{Y}\otimes SU(3)_{F}\otimes U(1)_{N}\otimes Z_{2}$, among them, the first three subgroups are namely the SM symmetry at the low-energy scale. The family symmetry at the high-energy scale is characterized by the subgroups $SU(3)_{F}\otimes U(1)_{N}$. The subgroup $Z_{2}$ is a discrete symmetry at the middle-energy scale. The model particle contents and their quantum numbers under the family symmetry subgroups are listed in the following. The low-energy fermions and Higgs scalar fields consist of
\begin{alignat}{1}
 & Q_{L}\sim(3,0)\,,\hspace{0.5cm} u_{R}\sim(3,0)\,,\hspace{0.5cm} d_{R}\sim(3,0)\,,\nonumber\\
 & L_{L}\sim(3,0)\,,\hspace{0.5cm} \nu_{R}\sim(3,0)\,,\hspace{0.5cm} e_{R}\sim(3,0)\,,\nonumber\\
 & H_{1}\sim(1,2)\,,\hspace{0.5cm} H_{2}\sim(1,-2)\,,\hspace{0.5cm} \phi\sim(1,0)\,.
\end{alignat}
 The three generation of fermions are in $\mathbf{3}$ representation of the family subgroup $SU(3)_{F}$, and they have no charges of the subgroup $U(1)_{N}$, and so on. Under the SM group, the representations of these fields are clear as usual but only $\nu_{R}$ and $\phi$ are singlets.

 We introduce some super-heavy fermion fields as follows
\begin{alignat}{1}
 &\eta^{u,\nu}_{1}\sim(3,2)\,,\hspace{0.3cm} \eta^{u,\nu}_{2}\sim(3,\frac{5}{3})\,,\hspace{0.3cm}
  \eta^{u,\nu}_{3}\sim(3,\frac{4}{3})\,,\hspace{0.3cm} \eta^{u,\nu}_{4}\sim(3,\frac{2}{3})\,,\nonumber\\
 &\eta^{u,\nu}_{5}\sim(3,\frac{3}{2})\,,\hspace{0.3cm} \eta^{u,\nu}_{6}\sim(3,1)\,,\hspace{0.3cm}
  \eta^{u,\nu}_{7}\sim(3,\frac{1}{2})\,,\hspace{0.3cm} \zeta^{u}\sim(1,1)\,,\hspace{0.3cm} \chi^{\nu}\sim(3,0)\,,\nonumber\\
 &\eta^{d,e}_{1}\sim(3,-2)\,,\hspace{0.3cm} \eta^{d,e}_{2}\sim(3,-\frac{5}{3})\,,\hspace{0.3cm}
  \eta^{d,e}_{3}\sim(3,-\frac{4}{3})\,,\hspace{0.3cm}\eta^{d,e}_{4}\sim(3,-\frac{2}{3})\,,\hspace{0.3cm}\zeta^{d,e}\sim(1,-1)\,,
\end{alignat}
 whose left-handed and right-handed fields are unified. The superscripts respectively indicate the corresponding right-handed fermions in (1), namely under the SM group the quantum numbers of the super-heavy fermions are the same as ones of the low-energy right-handed fermions. In comparison with the super-heavy quark fields, the super-heavy lepton fields have $\chi^{\nu}$ instead of $\zeta^{\nu}$. These super-heavy quarks and leptons are possessed of the super-heavy masses, so they appear only in the very high energy circumstances.

 We also introduce the super-heavy scalar flavon fields such as
\begin{alignat}{1}
 & F_{1}\sim(8,\frac{1}{3})\,,\hspace{0.5cm} F_{2}\sim(8,\frac{5}{3})\,,\hspace{0.5cm}
   F_{3}\sim(8,\frac{2}{3})\,,\hspace{0.5cm} F_{4}\sim(8,\frac{1}{2})\,,\nonumber\\
 & T_{1}\sim(3,1)\,,\hspace{0.5cm} T_{2}\sim(3,-1)\,.
\end{alignat}
 All of them are singlets under the SM group, but under the family subgroup, $F_{1},\cdots,F_{4}$ are hermitian octet representations, and $T_{1},T_{2}$ are complex triplet representations. In addition, they have different charges of $U(1)_{N}$. These flavon fields are responsible for the family symmetry breaking.

 Finally, we define the discrete group $Z_{2}$ as follows. Only the $\nu_{R}$ and $\phi$ fields are transformed as
\ba
 \nu_{R}\longrightarrow-\nu_{R}\,,\hspace{0.5cm} \phi\longrightarrow-\phi\,,
\ea
 and all of the other fields are uniformly transformed as themselves.

 Under the model symmetry group, the gauge invariant Yukawa couplings in the quark sector are written as
\begin{alignat}{1}
 \mathscr{L}_{q}=
 &\:\overline{Q_{L}}H_{2}\eta^{u}_{1R}+\overline{\eta^{u}_{1L}}\left(T_{1}\zeta^{u}_{R}+F_{1}\eta^{u}_{2R}+F_{3}\eta^{u}_{3R}
  +F_{4}\eta^{u}_{5R}\right)+\overline{\eta^{u}_{3L}}F_{3}\eta^{u}_{4R} \nonumber\\
 &+\overline{\eta^{u}_{5L}}F_{4}\eta^{u}_{6R}+\overline{\eta^{u}_{6L}}F_{4}\eta^{u}_{7R}+\left(\overline{\zeta^{u}_{L}}T^{\dag}_{2}
  +\overline{\eta^{u}_{2L}}F_{2}+\overline{\eta^{u}_{4L}}F_{3}+\overline{\eta^{u}_{7L}}F_{4}\right)u_{R} \nonumber\\
 &+\overline{Q_{L}}H_{1}\eta^{d}_{1R}+\overline{\eta^{d}_{1L}}\left(T_{2}\zeta^{d}_{R}+F^{*}_{1}\eta^{d}_{2R}+F^{*}_{3}\eta^{d}_{3R}\right)
  +\overline{\eta^{d}_{3L}}F^{*}_{3}\eta^{d}_{4R} \nonumber\\
 &+\left(\overline{\zeta^{d}_{L}}T^{\dag}_{1}+\overline{\eta^{d}_{2L}}F^{*}_{2}+\overline{\eta^{d}_{4L}}F^{*}_{3}\right)d_{R}
  +h.c.\,.
\end{alignat}
 For the sake of concision, we have left out the coupling coefficient at the front of each term in (5), which should be $\thicksim\mathscr{O}(1)$. Easy to notice, the couplings in the up-type sector are different from ones in the down-type sector. Those F-type flavon fields in the down-type sector are complex conjugate form. The Yukawa couplings in the lepton sector are similarly given as
\begin{alignat}{1}
 \mathscr{L}_{l}=
 &\:\overline{L_{L}}H_{2}\eta^{\nu}_{1R}+\overline{\eta^{\nu}_{1L}}\left(F_{1}\eta^{\nu}_{2R}+F_{3}\eta^{\nu}_{3R}
  +F_{4}\eta^{\nu}_{5R}\right)+\overline{\eta^{\nu}_{3L}}F_{3}\eta^{\nu}_{4R} \nonumber\\
 &+\overline{\eta^{\nu}_{5L}}F_{4}\eta^{\nu}_{6R}+\overline{\eta^{\nu}_{6L}}F_{4}\eta^{\nu}_{7R}+\left(\overline{\eta^{\nu}_{2L}}F_{2}
  +\overline{\eta^{\nu}_{4L}}F_{3}+\overline{\eta^{\nu}_{7L}}F_{4}\right)\chi^{\nu}_{R}+\overline{\chi^{\nu}_{L}}\phi\nu_{R}\nonumber\\
 &+\overline{L_{L}}H_{1}\eta^{e}_{1R}+\overline{\eta^{e}_{1L}}\left(T_{2}\zeta^{e}_{R}+F^{*}_{1}\eta^{e}_{2R}
  +F^{*}_{3}\eta^{e}_{3R}\right)+\overline{\eta^{e}_{3L}}F^{*}_{3}\eta^{e}_{4R} \nonumber\\
 &+\left(\overline{\zeta^{e}_{L}}T^{\dag}_{1}+\overline{\eta^{e}_{2L}}F^{*}_{2}+\overline{\eta^{e}_{4L}}F^{*}_{3}\right)e_{R}
  +h.c.\,,
\end{alignat}
 likewise, all the coupling coefficients are omitted. In comparison with the quark sector, the lepton sector has the exclusive terms related to $\chi^{\nu}$ instead of $\zeta^{\nu}$. These differences play key roles in generating distinct masses and mixings for quarks and leptons. The (5) and (6) Lagrangian indicate that the flavor interactions among the super-heavy fermions are transmitted by means of the super-heavy flavon fields, and the effects is ultimately transferred to the low-energy fermions after the multiple transmissions.

\vspace{1cm}
 \noindent\textbf{III. Symmetry Breakings and Fermion Mass Matrices}

\vspace{0.3cm}
 The model symmetry breakings go through three stages. The first step of the breaking chain is that the subgroups $SU(3)_{F}\otimes U(1)_{N}$ break to nothing, namely the family symmetry vanishes. This is implemented by the flavon fields $F_{1},\cdots,F_{4},T_{1},T_{2}$ developing VEVs along specified directions in the family space. The detailed vacuum structures are as follows
\begin{alignat}{1}
 & \frac{\langle F_{1}\rangle}{\Lambda_{F}}\sim\varepsilon\,\lambda_{6}\,,\hspace{0.5cm}
   \frac{\langle F_{2}\rangle}{\Lambda_{F}}\sim\frac{\varepsilon}{\sqrt{14}}\left(3\,\lambda_{3}
                                            -\sqrt{2}\,\lambda_{6}+\sqrt{3}\,\lambda_{8}\right)\,,\nonumber\\
 & \frac{\langle F_{3}\rangle}{\Lambda_{F}}\sim\frac{\varepsilon}{2}\left(\lambda_{1}+\lambda_{2}
                                            +\lambda_{6}-\lambda_{7}\right)\,,\hspace{0.5cm}
   \frac{\langle F_{4}\rangle}{\Lambda_{F}}\sim\varepsilon\,\lambda_{1}\,,\nonumber\\
 & \frac{\langle T_{1}\rangle}{\Lambda_{F}}\sim\left(\begin{array}{c}0\\0\\1\end{array}\right)\,,\hspace{0.5cm}
   \frac{\langle T_{2}\rangle}{\Lambda_{F}}\sim\left(\begin{array}{c}0\\0\\1\end{array}\right)\,,
\end{alignat}
 where $\Lambda_{F}$ is the family symmetry breaking scale, that is the dynamics scale of the super-heavy fermions and flavons, which is usually close to Planck scale of $10^{19}$ GeV. The only breaking parameter $\varepsilon$ is a ratio of the F-type VEV to the T-type VEV. We consider that the former is one order of magnitude smaller than the latter, thus $\varepsilon$ is $\thicksim \mathscr{O}(0.1)$. The matrices $\lambda_{1},\lambda_{2},\cdots,\lambda_{8}$ are the standard Gell-Mann matrices representing the generators of $SU(3)$. As before one coefficient of $\mathscr{O}(1)$ is implied in the right formula of each wave notation in (7). Below the scale $\Lambda_{F}$, all of the flavon fields develop the vacuum states with the structures, consequently the family symmetry is broken. It can be seen from (7) that the breakings of $T_{1}$ and $T_{2}$ bring the family symmetry down from $SU(3)$ to $SU(2)$. On the other hand, the breakings of $F_{1}$ and $F_{2}$ occur along the direction of the subgroup $S_{2}(2\leftrightarrow3)$ in the family space, which is a permutation group between the second generation fermions and the third ones, and the $F_{3}$ and $F_{4}$ breakings are respectively in the directions of the subgroup $S_{2}(1\leftrightarrow3)$ and $S_{2}(1\leftrightarrow2)$. The multiple breakings lead the family symmetry to disappear completely. This set of breaking patterns can be determined in principle by the self-interaction potential of every flavon field, but we do not go into the details. We directly adopt the intuitive and reasonable scenarios, in particular, it turns out to be a great success in fitting experimental data.

 When the energy scale goes down to the value far smaller than $\Lambda_{F}$, all the super-heavy fermions are actually decoupling. After all of them are integrated out from the original Lagrangian, then an effective Yukawa Lagrangian at the low energy is derived as
\ba
 \mathscr{L}^{eff}_{Yukawa}=\overline{Q_{L}}H_{2}Y_{u}u_{R}+\overline{Q_{L}}H_{1}Y_{d}d_{R}
                     +\overline{L_{L}}H_{1}Y_{e}e_{R}+\overline{L_{L}}H_{2}Y_{\nu}\frac{\phi}{\Lambda_{F}}\nu_{R}+h.c.
\ea
 with Yukawa coupling matrices
\begin{alignat}{1}
 &Y_{u}=y^{u}_{1}R_{1}+y^{u}_{2}\,\varepsilon^{2}R_{2}+y^{u}_{3}\,\varepsilon^{3}R_{3}+y^{u}_{4}\,\varepsilon^{4}R_{4}\,,\nonumber\\
 & Y_{d}=y^{d}_{1}R_{1}+y^{d}_{2}\,\varepsilon^{2}R_{2}+y^{d}_{3}\,\varepsilon^{3}R^{*}_{3}\,,\nonumber\\
 & Y_{e}=y^{e}_{1}R_{1}+y^{e}_{2}\,\varepsilon^{2}R_{2}+y^{e}_{3}\,\varepsilon^{3}R^{*}_{3}\,,\nonumber\\
 & Y_{\nu}=y^{\nu}_{1}\,\varepsilon^{2}R_{2}+y^{\nu}_{2}\,\varepsilon^{3}R_{3}+y^{\nu}_{3}\,\varepsilon^{4}R_{4}\,,
\end{alignat}
where
\begin{alignat}{1}
 & R_{1}=\left(\begin{array}{ccc}0&0&0\\0&0&0\\0&0&1\\\end{array}\right)\!,\hspace{0.5cm}
   R_{2}=\left(\begin{array}{ccc}0&0&0\\0&-1&-\sqrt{2}\\0&-\sqrt{2}&-1\\\end{array}\right)\!,\nonumber\\
 & R_{3}=\frac{1}{\sqrt{2}}\left(\begin{array}{ccc}0&1-i&0\\1+i&0&1+i\\0&1-i&0\\\end{array}\right)\!,\hspace{0.5cm}
   R_{4}=\left(\begin{array}{ccc}1&0&0\\0&1&0\\0&0&0\\\end{array}\right)\!.
\end{alignat}
 In (9), $y^{u}_{1},\cdots,y^{\nu}_{3}$ are some effective coupling coefficients, which are left out before, now they are visibly retrieved and written out. These coefficients are mostly $\thicksim \mathscr{O}(1)$, we take them as real numbers without loss of generality, so all the Yukawa matrices are hermitian. We can illustrate this procedure, for instance, the terms of $R_{1}$ and $R_{2}$ in the $Y_{u}$ matrix are generated respectively by (a) and (b) in the Figure 1, and so on.
 \begin{figure}
 \centering
 \includegraphics[totalheight=7cm]{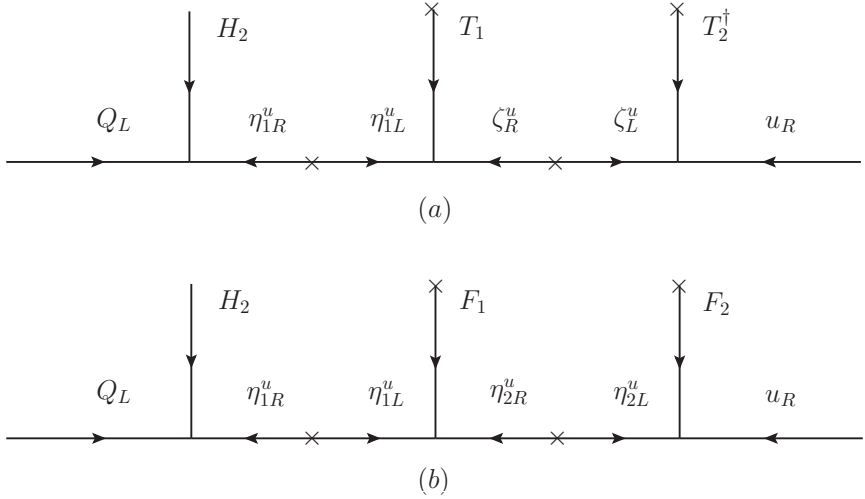}
 \caption{The graph of generating the effective Yukawa coupling matrices from the family symmetry breaking, (a) and (b) respectively give rise to the terms of $R_{1}$ and $R_{2}$ in $Y_{u}$\,.}
\end{figure}
 The effective theory, which is valid until the scale $\Lambda_{F}$, includes the SM fermions as well as three singlet right-handed neutrinos, moreover, has two doublet and one singlet Higgs fields. Since the term involving $\nu_{R}$ and $\phi$ in (8) is drastically suppressed by $\Lambda_{F}$, it is far smaller than the other terms in Lagrangian, furthermore, it is also non-renormalizable. It is very clear from (9) and (10) that this set of Yukawa coupling matrices indeed have some regular and intrinsic relations. Several notable characteristics can be seen very easy. First, every Yukawa matrix is a linear combination from the four fundamental matrices $R_{1},\cdots,R_{4}$, and the combination coefficients are expanded by a power series of $\varepsilon$. By virtue of such structures the elements of every Yukawa matrix show themselves large hierarchy. As a result, the $R_{1}$ and $R_{2}$ terms, namely the $\varepsilon^{0}$ and $\varepsilon^{2}$ terms, will respectively dominate the third and second generation fermion masses. The rest of the terms will make main contributions to the first generation fermion mass. Second, in contrast with $Y_{u},Y_{d},Y_{e}$, the leading term of $R_{1}$ is no in $Y_{\nu}$. In view of these structure features of the Yukawa matrices, it is in the course of nature that the transformation matrices diagonalizing $Y_{u},Y_{d},Y_{e}$ are all close to the unit matrix, whereas the transformation matrix diagonalizing $Y_{\nu}$ is approximately the tri-bimaximal mixing pattern. This difference is the principal source of generating distinct flavor mixings for the quarks and leptons. Third, the imaginary elements of $R_{3}$ are the only source of the $C$ and $CP$ violations in the Yukawa sector. To sum up, the four $R$-type matrices and the $\varepsilon$ parameter all together make up the skeleton frame of every Yukawa matrix, therefore they play key roles in fermion masses and flavor mixings.

 The second step of the breaking chain is that the discrete subgroup $Z_{2}$ is broken by the real singlet scalar field $\phi$ developing VEV as follows
\ba
 \frac{\langle \phi \rangle}{\Lambda_{F}}=\kappa\,.
\ea
 The breaking scale is considered as some intermediate value between the family breaking scale $\Lambda_{F}$ and the electroweak breaking scale. If the parameter $\kappa$ is about $10^{-10}$ or so, the effective Yukawa couplings of the neutrinos will be drastically suppressed owing to the $\kappa$ factor, thus they will be far smaller than the Yukawa couplings of the charged fermions. This possibly becomes a source of the neutrino tiny masses, of course, it is different from the usual see-saw mechanism \cite{19}.

 After the $Z_{2}$ breaking, the model remaining symmetry is exactly the SM symmetry group. The last step of the breaking chain is namely the electroweak symmetry breaking. It is accomplished by the doublet Higgs fields $H_{1}$ and $H_{2}$ developing VEVs as follows
\ba
 \frac{\langle H_{1}\rangle}{v_{ew}}=\left(\begin{array}{c}0\\cos\beta\end{array}\right),\hspace{0.5cm}
 \frac{\langle H_{2}\rangle}{v_{ew}}=\left(\begin{array}{c}sin\beta\\0\end{array}\right),
\ea
 where $v_{ew}$ is the electroweak scale and $tan\beta$ is a ratio of the up-type VEV to the down-type VEV. After the electroweak breaking all of the quarks and leptons obtain Dirac masses. The whole fermion mass terms are now given as
\ba
 -\mathscr{L}_{mass}=\overline{u_{L}}\,M_{u}u_{R}+\overline{d_{L}}\,M_{d}d_{R}
                     +\overline{e_{L}}\,M_{e}e_{R}+\overline{\nu_{L}}\,M_{\nu}\nu_{R}+h.c.
\ea
 with the mass matrices
\begin{alignat}{1}
 & M_{u}=-v_{ew}sin\beta\,Y_{u}\,,\hspace{0.5cm} M_{\nu}=-\kappa\,v_{ew}sin\beta\,Y_{\nu}\,,\nonumber\\
 & M_{d}=-v_{ew}cos\beta\,Y_{d}\,,\hspace{0.5cm} M_{e}=-v_{ew}cos\beta\,Y_{e}\,.
\end{alignat}
 Now three new parameters $\kappa,v_{ew},tan\beta$ are added into the model besides the $\varepsilon$ and y-type parameters in (9). All of these quantities are undetermined except the electroweak scale $v_{ew}$. However, some parameters among them are in the form of product factors in the mass matrices, there are actually three non-independent parameters. We have some freedoms to remove the non-independent parameters, for example, this three parameters $y^{u}_{2},y^{d}_{2},y^{\nu}_{1}$ can be absorbed collectively by redefinitions of the three parameters $\varepsilon,tan\beta,\kappa$, so each of them will be equal to one instead of free parameters hereinafter. Therefore, the effective independent parameters in the model are only thirteen in all. It can be seen from (14) that $v_{ew}$ dominates mass scale of the quarks and the charged leptons, $tan\beta$ is responsible for mass split of the up-type and down-type fermions, and the factor $\kappa$ causes that the neutrino masses are far smaller than ones of the charged fermions. Since this three parameters are only some product factors in the mass matrices, anyway, they have no influence on the flavor mixings. The mass hierarchy and the flavor mixings are still controlled mainly by the $\varepsilon$ parameter and the $R$-type matrices. In a word, this set of mass matrices properly embody all of the information about fermion mass hierarchy, flavor mixings and the $CP$ violations.

 In virtue of the model's intrinsic characteristics, all the fermion mass matrices are hermitian, therefore all of fermion mass eigenvalues are conveniently solved by diagonalizing them as follows
\begin{alignat}{1}
 & U_{u}^{\dagger}\,M_{u}\,U_{u}=\mathrm{diag}\left(m_{u},m_{c},m_{t}\right),\hspace{0.5cm}
   U_{\nu}^{\dagger}\,M_{\nu}\,U_{\nu}=\mathrm{diag}\left(m_{1},m_{2},m_{3}\right),\nonumber\\
 & U_{d}^{\dagger}\,M_{d}\,U_{d}=\mathrm{diag}\left(m_{d},m_{s},m_{b}\right),\hspace{0.5cm}
   U_{e}^{\dagger}\,M_{e}\,U_{e}=\mathrm{diag}\left(m_{e},m_{\mu},m_{\tau}\right).
\end{alignat}
 Because the $R_{1}$ and $R_{2}$ terms are respectively the leading and next-to-leading terms in the mass matrices, the second and third generation of the quark and charged lepton masses can be calculated approximately such as
 \begin{alignat}{1}
 & m_{c}\approx v_{ew}sin\beta\left(\varepsilon^{2}+\varepsilon^{4}\left(2+y^{u2}_{3}-y^{u}_{4}\right)\right)\,,\hspace{0.5cm}
   m_{t}\approx v_{ew}sin\beta\left(y^{u}_{1}-\varepsilon^{2}\right)\,,\nonumber\\
 & m_{s}\approx v_{ew}cos\beta
                      \left(\varepsilon^{2}+\varepsilon^{4}\left(\frac{2}{y^{d}_{1}}+y^{d2}_{3}\right)\right)\,,\hspace{0.5cm}
   m_{b}\approx v_{ew}cos\beta\left(-y^{d}_{1}+\varepsilon^{2}\right)\,,\nonumber\\
 & m_{\mu}\approx v_{ew}cos\beta
   \left(\varepsilon^{2}y^{e}_{2}+\varepsilon^{4}\left(\frac{2y^{e}_{2}}{y^{e}_{1}}+\frac{y^{e2}_{3}}{y^{e}_{2}}\right)\right)\,,\hspace{0.5cm}
   m_{\tau}\approx v_{ew}cos\beta\left(y^{e}_{1}-\varepsilon^{2}y^{e}_{2}\right)\,.
\end{alignat}
 However, the first generation of the quark and charged lepton masses have no such approximate expressions since they depend on all the terms of every mass matrix. It can be seen from (16) that, in the leading approximation, there are the mass relations
\ba
 \frac{m_{c}}{m_{t}}\approx \frac{\varepsilon^{2}}{y^{u}_{1}-\varepsilon^{2}}\,,\hspace{0.6cm}
 \frac{m_{s}}{m_{b}}\approx \frac{\varepsilon^{2}}{-y^{d}_{1}+\varepsilon^{2}}\,,\hspace{0.6cm}
 \frac{m_{\mu}}{m_{\tau}}\approx
                         \frac{\varepsilon^{2}}{\,\frac{y^{e}_{1}}{y^{e}_{2}}-\varepsilon^{2}}\,.
\ea
 We can easily estimate values of some parameters from (16) and (17). Finally, the flavor mixing matrices for the quarks and leptons are respectively given by \cite{20}
 \ba U_{u}^{\dagger}\,U_{d}=U_{CKM}\,,\hspace{0.5cm}
     U_{e}^{\dagger}\,U_{\nu}=U_{PMNS}\,.
 \ea
 The mixing angles and $CP$-violating phases in the two unitary matrices of $U_{CKM}$ and $U_{PMNS}$ can be worked out by the standard parameterization in particle data group \cite{1}.

\vspace{1cm}
 \noindent\textbf{IV. Numerical Results}

\vspace{0.3cm}
 Now we present the model numerical results. As is noted earlier, altogether the model parameters involve the ten y-type coefficients, the three breaking parameters $\varepsilon,\kappa,tan\beta$, and the electroweak scale $v_{ew}$. Once this set of parameters are chosen as the input values, according to the model we can calculate the various output values of the fermion masses and flavor mixings, moreover, all of the results can be compared with the current and future experimental data.

 The electroweak scale $v_{ew}$ is essentially determined in the gauge sector by weak gauge boson masses and gauge coupling constant. The accurate measures have given $v_{ew}=174$ GeV. The other thirteen parameters are really free parameters in the Yukawa sector, however, all of them have to be fixed by fitting the experimental data of the fermion masses and flavor mixings. Because the number of the model parameters is much less than the experimental values of the masses and mixings, and the majority of them have precisely been measured, the space of the model parameters is constrained very narrow and the tuning scope of the parameters is indeed very small. Although the fit is a non-trivial and no easy one, in advance one can find some parameter values by (16) and (17), and then the global fit can be successfully finished. We here give the values of the best fit instead of the detailed numerical analysis. The input values of the model parameters are elaborately chosen as follows
\begin{alignat}{1}
 & \varepsilon=0.0845\,,\hspace{0.5cm} \kappa=1.83\times10^{-11}\,,\hspace{0.5cm} tan\beta=12.37\,,\nonumber\\
 & y^{u}_{1}=1\,,\hspace{0.5cm} y^{u}_{3}=-1\,,\hspace{0.5cm} y^{u}_{4}=-0.7\,,\hspace{0.5cm}
   y^{d}_{1}=-0.292\,,\hspace{0.5cm} y^{d}_{3}=2.58\,,\nonumber\\
 & y^{\nu}_{2}=3.62\,,\hspace{0.5cm} y^{\nu}_{3}=7.3\,,\hspace{0.5cm}
   y^{e}_{1}=0.133\,,\hspace{0.5cm} y^{e}_{2}=0.96\,,\hspace{0.5cm} y^{e}_{3}=0.871\,.
\end{alignat}
 These values are reasonable and consistent with the previous estimates. Each of the y-type parameters is dedicated to a certain impact on the fermion masses and flavor mixings, for instance, $y^{u}_{3}$ has main impact on the quark $sin\theta^{q}_{13}$ and $CP$-violating phase, while $y^{d}_{3}$ makes main contribution to the quark $sin\theta^{q}_{12}$, and so forth.

 Finally, a variety of the numerical results calculated by the model are in detail listed in the following. For the quark sector, all of mass eigenvalues and mixing angles are (mass in GeV unit)
\begin{alignat}{1}
 & m_{u}=0.00246\,,\hspace{0.5cm} m_{c}=1.272\,,\hspace{0.5cm} m_{t}=172.2\,,\nonumber\\
 & m_{d}=0.00472\,,\hspace{0.5cm} m_{s}=0.101\,,\hspace{0.5cm} m_{b}=4.2\,, \nonumber\\
 & s^{\,q}_{12}=0.2255\,,\hspace{0.3cm} s^{\,q}_{23}=0.0414\,,\hspace{0.3cm} s^{\,q}_{13}=0.00342\,,\hspace{0.3cm}
   \delta^{\,q}=0.375\,\pi \approx 67.5^{\circ}\,,
\end{alignat}
 where $s_{\alpha\beta}=sin\theta_{\alpha\beta}$, in addition, the Jarlskog invariant measuring the $CP$ violation is calculated to
\ba
 J_{CP}^{\,q}\approx2.87\times10^{-5}\,.
\ea
 It is very clear that the above results are very well in agreement with the current measures of the quark masses, mixing and $CP$ violation \cite{1}. Although the first generation of the quark masses have not been accurately measured so far, their values are finely predicted to be about the center values of the experimental limits. For the lepton sector, the parallel results are
\begin{alignat}{1}
 &m_{e}=0.511\:\mathrm{MeV}\,,\hspace{0.5cm}m_{\mu}=105.7\:\mathrm{MeV}\,,\hspace{0.5cm}m_{\tau}=1778\:\mathrm{MeV}\,,\nonumber\\
 & m_{1}=0.172\times10^{-2}\:\mathrm{eV}\,,\hspace{0.5cm} m_{2}=0.888\times10^{-2}\:\mathrm{eV}\,,\hspace{0.5cm}
   m_{3}=5.01\times10^{-2}\:\mathrm{eV}\,,\nonumber\\
 & s^{\,l}_{12}=0.565\,,\hspace{0.3cm} s^{\,l}_{23}=0.751\,,\hspace{0.3cm} s^{\,l}_{13}=0.109\,,\hspace{0.3cm}
   \delta^{\,l}=-0.82\,\pi \approx -147.6^{\circ}\,.
\end{alignat}
 The charged lepton masses are completely identical with ones in the particle list \cite{1}. For the sake of comparison with the experimental data, the common used quantities in neutrino physics are explicitly calculated as follows
\begin{alignat}{1}
 & \triangle m^{2}_{21}\approx 7.60\times10^{-5}\:\mathrm{eV^{2}}\,,\hspace{0.5cm}
   \triangle m^{2}_{32}\approx 2.43\times10^{-3}\:\mathrm{eV^{2}}\,,\nonumber\\
 & sin^{2}2\theta^{\,l}_{12}\approx 0.869\,,\hspace{0.5cm} sin^{2}2\theta^{\,l}_{23}\approx 0.984\,,\hspace{0.5cm}
   sin^{2}2\theta^{\,l}_{13}\approx 0.047\,,\nonumber\\
 & J_{CP}^{\,l}\approx-0.01334\,,
\end{alignat}
 where $\triangle m^{2}_{\alpha\beta}=m^{2}_{\alpha}-m^{2}_{\beta}$. These results are excellently in agreement with the recent neutrino oscillation data \cite{21}. In particular, the model predicts that the heaviest one of the three generation neutrinos is about $0.05$ eV, the lepton mixing angle $\theta^{\,l}_{13}$ is $\thicksim3.3^{\circ}$ but nonzero, in addition, the $CP$-violating effect is of the order of $10^{-2}$ in the lepton sector, which is three order of magnitude larger than one in the quark sector. Since all the neutrinos in the model are Dirac-type rather than Majorana-type, the neutrinoless double beta decay is inevitably nought. Although these quantities have not strictly measured by now, some running and coming neutrino experiments are on the way toward these goals \cite{22}. We have confidence that all the predictions are promising to be tested in the near future.

 To sum up the above numerical results, in fact, only with the thirteen parameters does the model accurately and excellently fit the total twenty values of the fermion masses and flavor mixings. All the current measured values are exactly reproduced, meanwhile, all the non-detected values are finely predicted in the experimental limits. All of the results are naturally produced without any fine tuning. This fully show a strong prediction power of this model. In the case of the best fit, the sizes of the parameters $y^{u}_{1}$ and $y^{u}_{3}$ both are coincidently one, the reason about it is yet unknown and expected to research deeply.

\vspace{1cm}
 \noindent\textbf{V. Conclusions}

\vspace{0.3cm}
 In the paper, we have suggested a new model to solve the fermion masses and flavor mixings, which is based on the family symmetry $SU(3)_{F}\otimes U(1)_{N}$ and the discrete group $Z_{2}$. The family symmetry breaking is carried out by means of the introduced super-heavy fermion and flavon fields. After all of the super-heavy fermions are integrated out and decoupling, the low-energy effective theory is obtained with the regular Yukawa coupling matrices. The $\varepsilon_{0}$ parameter and the four fundamental $R$-type matrices all together make up the skeleton frame of the Yukawa matrices. In fact they play leading roles in the model, namely they dominate the fermion mass hierarchy and flavor mixing results. The discrete group $Z_{2}$ is broken by the singlet scalar field at the middle-energy scale. This leads that the Yukawa couplings of the neutrinos are drastically suppressed, and then gives rise to the tiny nature of the neutrino masses. That set of the fermion mass matrices derived from the model symmetries and their breakings are characterized only by the thirteen effective parameters. The model successfully and perfectly fits all the current experimental data about the fermion masses and flavor mixings, in particular, it finely predicts the first generation quark masses and the values of $\theta^{\,l}_{13}$, $J_{CP}^{\,l}$ in neutrino physics. All of the results are excellent and inspiring, and also fully show a great prediction power of the model. Finally, we expect all the results to be tested in future experiments on the ground and in the sky. These experiments will undoubtedly provide us more important information about the flavor physics, and then enlighten us to understand finely the mystery of the universe.

\vspace{1cm}
 \noindent\textbf{Acknowledgments}

\vspace{0.3cm}
 One of the authors, W. M. Yang, would like to thank his wife for large helps. This research is supported by chinese universities scientific fund.


\vspace{1cm}
\begin{thebibliography}{99}
\bibitem{1}
 C. Amsler \emph{et al.} [Particle Data Group], Phys. Lett. B 667, 1 (2008);
 K.Nakamura \emph{et al.} [Particle Data Group], J. Phys. G 37, 075021 (2010).
\bibitem{2}
 G. Altarelli, M. W. Grunewald, Phys. Reps. 403-404, 189 (2004).
\bibitem{3}
 A. Hocker and Z. Ligeti, Annu. Rev. Nucl. Part. Sci. 56, 501 (2006);
 R. Fleischer, arXiv:hep-ph/0608010;
 L. Camilleri, E. Lisi, and J. F. Wilkerson, Annu. Rev. Nucl. Part. Sci. 58, 343 (2008);
 R. N. Mohapatra, \emph{et al.}, Rep. Prog. Phys. 70, 1757 (2007).
\bibitem{4}
 Y. Fukuda \emph{et al.} [Super-Kamiokande Collaboration], Phys. Rev. Lett. 81, 1562 (1998); Phys. Rev. Lett. 85, 3999 (2000);
 M. Apollonio \emph{et al.} [CHOOZ Collaboration], Phys. Lett. B 466, 415 (1999); Eur. Phys. J. C 27, 331 (2003);
 K. Eguchi \emph{et al.} [KamLAND Collaboration], Phys. Rev. Lett. 90, 021802 (2003);
 Q. R. Ahmad \emph{et al.} [SNO Collaboration], Phys. Rev. Lett. 89, 011301 (2002).
\bibitem{5}
 F. T. Avignone III, S. R. Elliott, J. Engel, Rev. Mod. Phys. 80, 481 (2008).
\bibitem{6}
 M. Bona, \emph{et al.}[UTfit Collaboration], JHEP 0803, 049 (2008).
\bibitem{7}
 P. F. Harrison, D. H. Perkins and W. G. Scott, Phys. Lett. B 530, 167 (2002).
\bibitem{8}
 H. Nunokawa, S. Parke and J. Valle, Prog. Part. Nucl. Phys. 60, 338 (2008).
\bibitem{9}
 H. Fritzsch and Z. Z. Xing, Prog. Part. Nucl. Phys. 45, 1 (2000);
 R. N. Mohapatra and A. Y. Smirnov, Annu. Rev. Nucl. Part. Sci. 56, 569 (2006);
 H. Fritzsch, Int. J. Mod. Phys. A 24, 3354 (2009);
 K. S. Babu, arXiv:0910.2948.
\bibitem{10}
 M. Dine and A. Kusenko, Rev. Mod. Phys. 76, 1 (2004);
 W. Buchmuller, R. D. Peccei and T. Yanagida, Annu. Rev. Nucl. Part. Sci. 55, 311 (2005).
\bibitem{11}
 S. Davidson, E. Nardi and Y. Nir, Phys. Reps. 466, 105 (2008);
 E. Aprile and S. Profumo, New J. Phys. 11, 105002 (2009);
 K. Jedamzik and M. Pospelov, New J. Phys. 11, 105028 (2009).
\bibitem{12}
 G. Altarelli, arXiv:hep-ph/0610164;
 C. H. Albright and M. C. Chen, arXiv:hep-ph/0608137;
 G. Ross and M. Serna, arXiv:0704.1248.
\bibitem{13}
 C. D. Froggatt and H. B. Nielsen, Nucl. Phys. B 147, 277 (1979).
\bibitem{14}
 G. Altarelli and F. Feruglio, Nucl. Phys. B741, 215 (2006).
\bibitem{15}
 S. F. King and G. G. Ross, Phys.Lett. B 574, 239 (2003);
 G. G. Ross and L. Velasco-Sevilla, Nucl. Phys. B692, 50 (2004);
 I. M. Varzielas and G. G. Ross, Nucl. Phys. B733, 31 (2006);
 S. Antusch, S. F. King, M. Malinsky, JHEP 0806, 068 (2008).
\bibitem{16}
 W. M. Yang and H. H. Liu, Nucl. Phys. B820, 364 (2009).
\bibitem{17}
 M. C. Chen and K. T. Mahanthappa, Int. J. Mod. Phys. A 18, 5819 (2003);
 G. L. Kane, S. F. King, I. N. R. Peddie and L. V. Sevilla, JHEP 08, 083 (2005);
 S. F. King, JHEP 08, 105 (2005);
 C. Hagedorn, M. Lindner and R. N. Mohaptra, JHEP 06, 042 (2006);
 W. M. Yang and Z. G. Wang, Nucl. Phys. B707, 87 (2005).
\bibitem{18}
 J. C. Pati and A. Salam, Phys. Rev. D 10, 275 (1974);
 R. N. Mohapatra and G. Senjanovi\'{c}, Phys. Rev. D 23, 165 (1981).
\bibitem{19}
 M. Gell-Mann, P. Ramond, R. Slansky, in \emph{Supergravity}, eds. P. van Niewenhuizen and D. Z. Freeman
 (North-Holland, Amsterdam, 1979);
 T. Yanagida, in \emph{Proc. of the Workshop on Unified Theory and Baryon Number in the Universe},
 eds. O. Sawada and A. Sugamoto (Tsukuba, Japan, 1979);
 R. N. Mohapatra, G. Senjanovi\'{c}, Phys. Rev. Lett. 44, 912 (1980).
\bibitem{20}
 M. Kobayashi and T. Maskawa, Prog. Theor. Phys. 49, 652 (1973);
 B. M. Pontecorvo, Sov. Phys. JETP 6, 429 (1958);
 Z. Maki, M. Nakagawa and S. Sakata, Prog. Theor. Phys. 28, 870 (1962).
\bibitem{21}
 M. C. G.-Garcia, M. Maltoni, Phys. Reps. 460, 1 (2008);
 T. Schwetz, M. Tortola and J. Valle, New J. Phys. 10, 113011 (2008);
 G. L. Fogli, E. Lisi, A. Marrone, A. Palazzo, Pro. Part. Nucl. Phy. 57, 742 (2006);
 E. Lisi, Pro. Part. Nucl. Phy. 64, 171 (2010).
\bibitem{22}
 M. Nakahata, Nucl. Phys. Proc. Suppl. 145, 23 (2005);
 Ardellier F, et al. (Double Chooz Collab.) hep-ex/0606025v4 (2006);
 Guo X, et al. (Daya Bay Collab.) hep-ex/0701029v1 (2007);
 RENO Collab. http://neutrino.snu.ac.kr/RENO;
 Anjos JC, et al. Nucl. Phys. B Proc. Suppl. 155, 231 (2006);
 Hayato Y, et al. T2K Letter of Intent. http://jnusrv01.kek.jp/public/t2k/ (2003);
 Ayres DS, et al. (NOvA Collab.) hep-ex/0503053 (2005), http://www-nova.fnal.gov (2005);
 C. Arnaboldi, et al., Phys. Rev. Lett. 95, 142501 (2005);
 R. Arnold, et al., NEMO Collaboration, Nucl. Phys. A 781, 209 (2007).
\end{thebibliography}
\end{document}